# A Two-Parameter Model for Ultrasonic Tissue Characterization with Harmonic Imaging

*Kajoli Banerjee Krishnan[1], Nithin Nagaraj[2], Nitin Singhal[3], Shalini Thapar[4], Komal Yadav[4]*

[1]GE Global Research Centre, Bengaluru, India. [2]National Institute of Advanced Studies, Bengaluru, India. [3]Samsung Electronics, Bengaluru, India. [4]Institute of Liver and Biliary Sciences, Delhi, India.

Abstract

Over the past few decades, researchers have developed several approaches such as the Reference Phantom Method (RPM) to estimate ultrasound attenuation coefficient (AC) and backscatter coefficient (BSC). AC and BSC can help to discriminate pathology from normal tissue during in-vivo imaging. In this paper, we propose a new RPM model to simultaneously compute AC and BSC for harmonic imaging and a normalized score that combines the two parameters as a measure of disease progression. The model utilizes the spectral difference between two regions of interest, the first, a proximal, close to the probe and second, a distal, away from the probe. We have implemented an algorithm based on the model and shown that it provides accurate and stable estimates to within 5% of AC and BSC for simulated received echo from post-focal depths of a homogeneous liver-like medium. For practical applications with time gain and time frequency compensated in-phase and quadrature (IQ) data from ultrasound scanner, the method has been approximated and generalized to estimate AC and BSC for tissue layer underlying a more attenuative subcutaneous layer. The angular spectrum approach for ultrasound propagation in biological tissue is employed as a virtual Reference Phantom (VRP). The VRP is calibrated with a fixed probe and scanning protocol for application to liver tissue. In a feasibility study with 16 subjects, the method is able to separate 9/11 cases of progressive non-alcoholic fatty liver disease from 5 normal. In particular, it is able to separate 4/5 cases of non-alcoholic steato-hepatitis and early fibrosis (F≤2) from normal tissue. More extensive clinical studies are needed to assess the full capability of this model for screening and monitoring disease progression in liver and other tissues.

1. Introduction

Ultrasound attenuation coefficient (AC) and backscatter coefficient (BSC) of tissue have been of considerable interest in ultrasonic tissue characterization.[1-5] Two of the well-known frequency domain methods for computing AC are the spectral shift method and the spectral difference method both of which were first proposed, their assumptions and relative merits compared by Kuc.[6]

The Spectral shift method is premised on a Gaussian shaped radio-frequency (RF) envelope and measures AC from an estimation of the downward shift in the centroid of the RF pulse due to frequency dependent attenuation of ultrasound in tissue without explicitly accounting for effects of diffraction.[7] Statistical performance of this method compared to a parametric model of a power spectrum that incorporates effects of scan parameters, beam shape and attenuation in a tissue mimicking (TM) phantom shows that choice of optimal scan parameters can help to maximize the accuracy of its AC estimates over the imaging depth.[8]

The spectral difference method is based on computing AC from the slope representing the decay of the echo signal over an imaging path length in both narrow and broad band implementations.

[9, 10] Yao et al.[11] proposed a Reference Phantom Method (RPM) with known AC and BSC to account for ultrasound probe, diffraction and other depth dependent effects in the spectral difference method to estimate BSC of an unknown Sample. The method employs a ratio of the signal intensities of the Sample and Reference in a single region of interest (ROI). RPM has also been employed to calculate AC as the slope of the line that fits the log ratio of the spectral difference between a proximal and distal region of interest assuming a linear variation of attenuation with frequency.[12, 13]

Unlike the data reduction method proposed by Madsen et al.[14] RPM does not require an explicit model or measured beam pattern to estimate backscatter. RPM involves time domain processing, and assumes the same sound speed in the Reference and the Sample, no abrupt boundaries in the Sample and homogeneity in AC and BSC within the region of interest. Though not explicitly stated, its mathematical formulation assumes linear response of tissue particles to ultrasound (i.e. fundamental imaging).

The level of backscatter from any region of interest in tissue is affected by attenuation over the path that the ultrasound beam traverses preceding the depth of that region. Insana et al.[15] applied a modified Sigelmann-Reid technique to characterize BSC of tissue-like material by using a reference material with a known BSC and found it to be approximately proportional to particle size. Sleefe et al.[16] proposed a model for scatterer number density based on envelope statistics. The model took into consideration the probe and its interaction with tissue, and showed very good agreement with phantom experiments and tissue histology. Lizzi et al.[17] formulated three spectral characteristics: scatterer size, concentration and impedance in terms of attenuation and applied them in liver tissue characterization. Oelze et al.[18] cast the independent computation of average scatterer size and concentration as solution to an inverse problem from a mathematical expression for an attenuation-compensated gated echo signal.

More recently, to enable simultaneous computation of AC and BSC in layered Samples, Nam[19, 20] solved a least squares formulation of spectral difference RPM with constraints on AC and BSC using RF echo at three different frequencies and demonstrated its superior performance in characterizing layered TM phantoms with abrupt boundaries compared to the original RPM.

Several of these and other methods for estimating BSC and their subsequent refinements have been reviewed by Ghoshal et al.[21]

In this paper, we propose a new two-parameter model to make simultaneous estimation of AC and BSC with harmonic imaging and combine the parameters into a single normalized score. Our motivation stems from both practical and theoretical considerations. Harmonic is the default imaging mode for liver in most commercial ultrasound scanners, and hence, a method that could enable tissue characterization with a standard B-mode scan could be quite useful. Wallace et al.[22] proposed to measure nonlinearity ($\beta$) of a tissue type medium using ultrasonography and a referencing method without invoking the medium attenuation coefficient by using an expression for $\beta$ that is proportional to medium density and speed of sound in the medium. Increase in these parameters in disease states could then imply an increase in $\beta$. For example, increased density and sound speed has been reported for fibrotic liver.[23] Lin et al.[24] recently showed that harmonic backscatter echo modeled via Nakagami distribution can be associated with parameters that reveal characteristics such as scatterer distribution and its nonlinear behavior. If diseased tissue is indeed characterized by higher nonlinearity, and hence could led to higher harmonic generation and hence backscatter, harmonic imaging could utilize the mechanism as an additional discriminator between normal and diseased tissue.

Our basic model is an adaptation of the spectral difference method. It employs two ROI's with RPM as in Insana.[14] However, unlike Insana[14] where reflected signals from several moving pairs of ROI segments are used to estimate AC, in our approach one ROI is fixed and placed in a proximal location very close to the probe. The other distal ROI is placed beyond the focus. AC and BSC are evaluated for depths associated with the distal ROI. Like Yao et al.[11], we use the time domain echo signal. However, we extract the harmonic band from the signal for the purposes of analysis. The basic model in described in Section 2.

The model has been generalized for use in practical situations with in-phase and quadrature (IQ) signal from ultrasound scanners in in-vivo scans to (1) account for depth and frequency gain compensation that are embedded in the signal and (2) correct for the propagative effects of subcutaneous layer of tissue (3) obviate the need for an actual reference phantom scan by employing a virtual reference phantom (VRP). The generalized model is described in Section 3.

We demonstrate the ability of the basic model to extract AC and BSC from simulated ultrasound echoes in Section 4. A feasibility study for liver tissue characterization and its results on a 16-subject cohort is presented in Section 5. We summarize our findings in the concluding section 6.

## 2. Basic Model

In this section, we first recapitulate the spectral difference method of Yao et al.[11] and follow it with a presentation of our proposed adaptation of the method for harmonic imaging.

### 2.1 The Spectral Difference Method with RPM

Estimation of AC and BSC of a test Sample requires the compensation of effects of the transducer settings and sensitivity, and, diffraction related to transducer geometry. As illustrated in Figure 1, Yao et al.[11] had proposed a novel method of eliminating these effects, by using another scan with the same transducer settings on a Reference phantom whose acoustic properties (speed of sound, AC and BSC) are known. Assuming (a) average speed of sound in the Reference same as that in the Sample, (b) homogeneous Sample with isotropic scattering, (c) same diffraction due to transducer geometry in the Sample and Reference; AC and BSC can be estimated as follows.

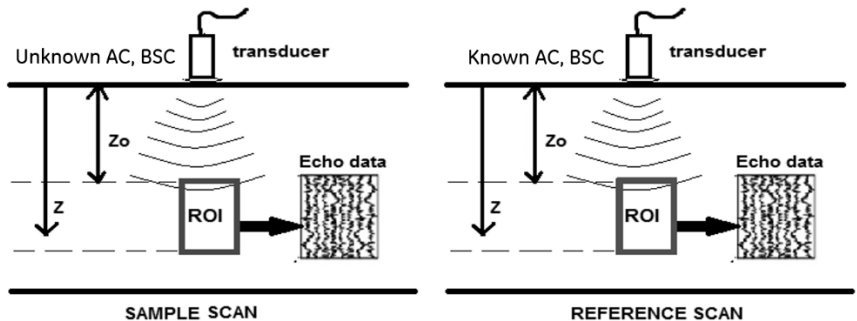

Figure 1. Reference Phantom Method.[11] Conventional RPM employs 1 ROI.

If $I_S(\omega_0, t)$ and $I_R(\omega_0, t)$ be the time-domain received echo intensity, band-pass filtered at frequency = $\omega_0$, then, their ratio can be expressed as:

$$\frac{I_S(\omega_0,t)}{I_R(\omega_0,t)} = \frac{BSC_S(\omega_0)\exp(-4\alpha_S\omega_0 z)}{BSC_R(\omega_0)\exp(-4\alpha_R\omega_0 z)} = \frac{I_S(\omega_0,z)}{I_R(\omega_0,z)} \qquad (1)$$

where the signal at time *t* is mapped to a depth of *z=c.t/2* where *c* = speed of sound. The subscript 'S' and 'R' stand for Sample and Reference respectively. Further, it is also assumed that attenuation is a linear function of frequency. In the above expression, $\alpha_S$ and $\alpha_R$ are expressed in Neper/cm-MHz.

The algorithm for determining AC and BSC from equation (1) is to simplify it further:

$$20 \log_{10}\left[\frac{I_S(\omega_0,z)}{I_R(\omega_0,z)}\right] = -4\omega_0 z(\Delta\alpha) + 20 \log_{10}\left[\frac{BSC_S(\omega_0)}{BSC_R(\omega_0)}\right] \qquad (2)$$

where $\Delta\alpha = \alpha_S - \alpha_R$ is expressed in dB/cm-MHz.

The absolute value of the intensity ratio is used in the computation of the left-hand side (LHS). From equation (2), if the best straight line is fitted between LHS and *z* over a ROI as shown in Figure 1, then:

$$\alpha_R - \alpha_S = \frac{Slope}{4\omega_0} \qquad (3)$$

$$\frac{BSC_S(\omega_0)}{BSC_R(\omega_0)} = 10^{\frac{Intercept}{20}} \qquad (4)$$

Thus, from equations (3) and (4), one can determine the AC and BSC of the Sample if we know the AC and BSC of the Reference phantom. Yao et al.[11] chose z to lie between 1–5 cm at frequencies ranging from 2.5 – 5 MHz.

Assumptions in this method imply that to enable comparison of equivalent ROI's (at the same depth and accounted for diffraction), the Reference must have the same sound speed as the Sample. Since the sound speed of the Sample is not known apriori, it is generally not possible to meet the needs of this assumption in practice. For example, sound speeds in liver have been reported to range from 1540 – 1650 m/s.[23] Mismatch of depths at the far ends of this range is approximately 3 mm at a ROI located at 4 cm, 6 mm for a ROI located at 8 cm and 9 mm for a ROI located at 16 cm depth. With a Reference phantom sound speed set at ~1570 m/s, the mismatches will be of much less than 1 cm across depths up to a depth of 16 cm. So, the errors likely to accrue with a ROI ~ 1 cm in an otherwise homogeneous Sample phantom are likely to be low. In a clinical setting, other effects such as inherent tissue heterogeneity and breathing motion are likely to cause greater variation in estimates than what is likely to arise out of mismatch in sound speed of this order.

## 2.2 A New Two-Parameter Model for Harmonic Imaging

One important implicit assumption in the formulation of Yao et al.[11] is that of linearity; both the Reference and the Sample are not only homogeneous and isotropic but also linear elastic medium, the received echo spectrum has the same frequencies as the transmitted spectrum. It assumes fundamental ultrasound imaging; not surprising since the work predated harmonic imaging in ultrasound systems.

Harmonic imaging is based on the nonlinear propagative effects of finite amplitude ultrasound. Nonlinear distortion of finite amplitude ultrasound in tissue generates acoustic signal at the second harmonic $\omega_0$ for a transmit frequency of $\omega_0/2$. Harmonic mode creates images from a finite bandwidth centered at $\omega_0$. The second harmonic has narrower main lobe and lower side lobes compared to the fundamental echo signal (at $\omega_0/2$).[25] It produces images of higher resolution without loss of penetration. The current generation ultrasound transducers have sufficient bandwidth to accommodate harmonics; and hence, harmonic is practically the default mode in most applications on commercial Ultrasound scanners.

The transmit pulse has little to no second harmonic near the transducer surface as it enters the tissue. Nonlinear distortion is an accumulative process; the second harmonic slowly builds up as the pulse propagates through the tissue. There are two other competitive phenomena that influence the growth of the second harmonic. These are (a) the position of the focus (for a focused pulse) and (b) the attenuation coefficient (AC) of the medium. The magnitude of the second harmonic is proportion to $P^2$ where P is the on-axis pressure magnitude of the fundamental beam.[26] Since the intensity of the transmit beam is highest in the focal region, the second harmonic tends to peak near the focus irrespective of the tissue (Reference or Sample). Frequency dependent attenuation on the other hand leads to a decay in the second harmonic and tends to dominate the propagation at large depths of tissue in the region beyond the focus. Thus, if the Reference and the Sample have different nonlinearity and attenuation, the ratio of their second harmonics could vary significantly over the propagation path *z*, and equation (1) may not hold. The ratio of the echo intensity for harmonic imaging may be then expressed as:

$$\frac{I_S(\omega_0,z)}{I_R(\omega_0,z)} = \frac{L_S(\omega_0,z)}{L_R(\omega_0,z)} \frac{BSC_S(\omega_0)\exp(-4\alpha_S\omega_0 z)}{BSC_R(\omega_0)\exp(-4\alpha_R\omega_0 z)} \tag{5}$$

The additional $L_S$ and $L_R$ are *lumped* terms that account for any other depth and frequency dependent variation such as consequences of nonlinear propagation that are not canceled by referencing.

Yao et al.[11] suggested the use of a sufficiently long duration time window to achieve a narrow band analysis of the signal to estimate AC and BSC with reasonable accuracy. They used a 4 μs window and demonstrated reasonable estimates with TM phantoms. They also cautioned against BSC estimation from short pulses; those can be prone to as much as 100% error if frequency dependence of BSC is unaccounted for when the AC and BSC are very different in the Sample and the Reference. Thus, *L* may also include depth dependence arising out of the nature of a relatively wider band harmonic IQ data that may not conform to the depth independence of backscatter assumed in Yao's formulation while performing estimates in the time domain.

ROI location could also influence accuracy of AC and BSC estimates. Yao et al.[11] recorded signals from depths ranging from 1-5 cm; Lu et al.[3] spanned a range of 2.3-12 cm. It has been reported that bias in attenuation estimation due to mismatch in sound speed between Reference and Sample is the worst at depths near the focal region and the least when the transmit focus is generally deep and deeper than the position of the ROI.[20]

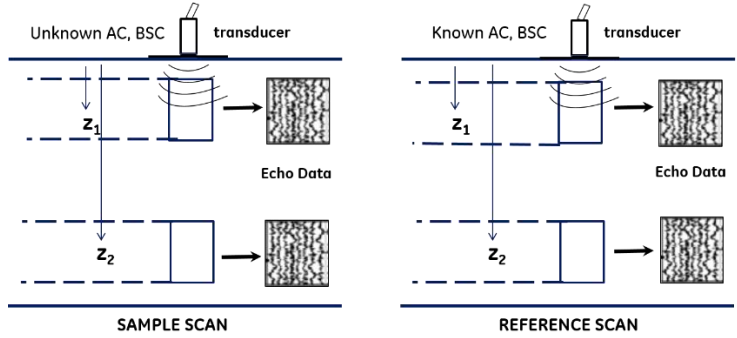

Figure 2. Reference Phantom Method for Harmonic Imaging employs 2 ROI's.

In case of harmonic imaging, we hypothesize the need to employ two ROI's to estimate both AC and BSC in a nonlinear homogeneous medium as illustrated in Figure 2. Thus, based on equation (5) and Figure 2, we write the ratio of the difference between echo intensity in the two ROI's as:

$$\frac{I_S(\omega_0,z_2)-I_S(\omega_0,z_1)}{I_R(\omega_0,z_2)-I_R(\omega_0,z_1)} = \frac{BSC_S(\omega_0)\exp(-4\alpha_S\omega_0 z_1)}{BSC_R(\omega_0)\exp(-4\alpha_R\omega_0 z_1)} \left[\frac{L_S(\omega_0,z_1)}{L_R(\omega_0,z_1)}\right] \left[\frac{\frac{L_S(\omega_0,z_2)}{L_S(\omega_0,z_1)}\exp(-4\alpha_S\omega_0 z_0)-1}{\frac{L_R(\omega_0,z_2)}{L_R(\omega_0,z_1)}\exp(-4\alpha_R\omega_0 z_0)-1}\right] \quad (6)$$

$$F(\omega_0,\alpha_S,\alpha_R,z_1,z_2) = \left[\frac{L_S(\omega_0,z_1)}{L_R(\omega_0,z_1)}\right] \left[\frac{\frac{L_S(\omega_0,z_2)}{L_S(\omega_0,z_1)}\exp(-4\alpha_S\omega_0 z_0)-1}{\frac{L_R(\omega_0,z_2)}{L_R(\omega_0,z_1)}\exp(-4\alpha_R\omega_0 z_0)-1}\right] \quad (7)$$

In equation (6), $z_1$ is the depth of the proximal ROI, $z_2$ is the depth of the distal ROI from the transducer and $z_2 = z_1 + z_0$. In comparison with equation (5) where there was only one variable z (location of a single ROI), equation (6) has two degrees of freedom namely $z_1$ and $z_2$ (location of the two ROIs). By appropriately choosing these two variables, we try to ensure that $F \cong 1$.

At the proximal ROI close to the transducer, the ultrasound beam has just entered the tissue. Propagation has barely been influenced by either attenuation or nonlinearity. So, $L_S(\omega_0, z_1)$ and $L_R(\omega_0, z_1) \cong 1$. The effect of nonlinearity dominates in the focal region; sound speed mismatch and finite bandwidth could also reduce the reliability of estimates from the focal region. A second distal ROI placed beyond the focus, a region in which attenuation plays a more significant role than nonlinearity, $L_S(\omega_0, z_2)$ and $L_R(\omega_0, z_2) \cong 1$. For typical values of transmit focus $z_0$ > 1 cm. For harmonic frequency ~ 4 MHz used in liver imaging, values of $\alpha_S$ reported in literature[27-29] and choice of $\alpha_R$ of the Reference Phantom, $\exp(-4\alpha_S\omega_0 z_0)$ and $\exp(-4\alpha_R\omega_0 z_0) \ll 1$. Therefore for a proximal ROI close to the transducer and distal ROI located beyond the focus F → 1 and equation (6) reduces to:

$$\frac{I_S(\omega_0,z_2)-I_S(\omega_0,z_1)}{I_R(\omega_0,z_2)-I_R(\omega_0,z_1)} = \frac{BSC_S(\omega_0)\exp(-4\alpha_S\omega_0 z_1)}{BSC_R(\omega_0)\exp(-4\alpha_R\omega_0 z_1)} \quad (8)$$

When this is achieved, we apply linear regression in the log-transformed equation (8). It then follows that AC can be estimated as the slope and BSC as the intercept of the linear fit. The absolute value of the intensity difference ratio is used in the computation of the LHS. In Section 4 we will show the performance of a Matlab implementation of equation 8 based on simulated ultrasound echo data.

## 3. Generalized Model

In this section, we present a practical extension of the basic model to account for pre-applied gain compensation in the IQ signal from an ultrasound scanner and a strategy to correct for propagative effects through the subcutaneous layer of tissue distinctive from the underlying tissue we intend to characterize. We also describe the formulation of a VRP.

3.1 Extension of Basic Model to Practice

We propose the following form for the LHS of equation (8) for in-vivo application in which the ultrasound beam typically encounters a subcutaneous tissue layer before entering the tissue of interest.

$$\frac{I_S(\omega_0,z_2)-I_S(\omega_0,z_1)}{I_R(\omega_0,z_2)-I_R(\omega_0,z_1)} = \frac{BSC_S(\omega_0)}{BSC_R(\omega_0)}\left[\frac{G_S(\omega_0,z_2)\exp(-4\alpha_f\omega_0\,\delta)\exp(4\alpha_S\omega_0\delta)\,exp(-4\alpha_S\omega_0 z_2)-G_S(\omega_0,z_1)exp(-4\alpha_f\omega_0 z_1)}{G_R(\omega_0,z_2)\,exp(-4\alpha_R\omega_0 z_2)-G_R(\omega_0,z_1)exp(-4\alpha_R\omega_0 z_1)}\right] \quad (9)$$

In addition to the expression in equation (1), $I_S$ and $I_R$, the received echo intensities from the Sample and Reference tissue respectively now have two more terms. The first of these is $G_S$ and $G_R$; *lumped* terms that primarily account for depth and frequency gain compensation that may be already applied to the echo signal from a commercial scanner prior to its usage in the computation of AC and BSC. The second is the attenuation of the ultrasound beam through the subcutaneous layer of thickness $\delta$ in the Sample tissue where $\alpha_f$ (in units of Neper/cm) is the attenuation coefficient of the layer. The backscatter coefficient of the subcutaneous layer is assumed to be $BSC_S$, the same as that of the Sample tissue. All other assumptions of the RPM and descriptions of the basic model described in section 2 hold.

Equation (9) may be rewritten by scaling the $I_S$ term at $z_2$ by compensating for the excess attenuation $\alpha_f - \alpha_R$ with respect to the attenuation of the Reference tissue through $\delta$, pulling out the first term in the numerator and the denominator from within the parentheses on the right-hand side (RHS) and defining a term H as the remaining multiplicative factor.

$$\frac{I_S(\omega_0,z_2)\exp(4(\alpha_f-\alpha_R)\,\omega_0\delta)-I_S(\omega_0,z_1)}{I_R(\omega_0,z_2)-I_R(\omega_0,z_1)} = \frac{BSC_S(\omega_0)G_S(\omega_0,z_2)}{BSC_R(\omega_0)G_R(\omega_0,z_2)}\left[\frac{exp(-4\alpha_S\omega_0\,(z_2-\delta))}{exp(-4\alpha_R\omega_0(z_2-\delta))}\right].H \quad (10)$$

$$H = \left[\frac{1-(G_S(\omega_0,z_1)/G_S(\omega_0,z_2))exp(4\alpha_S\omega_0\,(z_2-\delta)\,)exp(-4\alpha_f\omega_0 z_1)exp(4\alpha_R\omega_0\delta)}{1-(G_R(\omega_0,z_1)/G_R(\omega_0,z_2))exp(4\alpha_R\omega_0 z_0\,)}\right] \quad (11)$$

$G_s$ and $G_R$ are only system-dependent and not tissue-dependent variables. Hence, they are equal at same depths and frequency. H can be approximated to 1 when $\alpha_f$ and $\alpha_S$ are not very different from $\alpha_R$ and equation (11) can be log-transformed to:

$$20 \log_{10}\left[\frac{I_S(\omega_0,z_2)\exp(4(\alpha_f-\alpha_R)\omega_0\delta)-I_S(\omega_0,z_1)}{I_R(\omega_0,z_2)-I_R(\omega_0,z_1)}\right] = -4\omega_0(\Delta\alpha)(z_2-\delta) + 20\log_{10}\left[\frac{BSC_S(\omega_0)}{BSC_R(\omega_0)}\right]$$

(12)

$\Delta\alpha = \alpha_S - \alpha_R$ is expressed in dB/cm/MHz. $\alpha_s$ can be estimated from the slope of the line fitted over the path to the distal ROI w.r.t the end of the subcutaneous layer. $BSC_s$ can be estimated from the intercept of the fitted line.

The approximation for H is also expected to hold when $\alpha_f > \alpha_R$ since the proximal ROI depth $z_1$ is small. However, errors could accrue at larger distal depths $z_2$ when $\alpha_S > \alpha_R$ with increase in the thickness of the subcutaneous layer $\delta$.

3.2 Virtual Reference Phantom (VRP)

A single line of echo data can be simulated using an angular spectrum method to account for diffraction, frequency dependent attenuation and nonlinear distortion of finite amplitude ultrasound in tissue transmitted and received by a circular aperture.[26,27,30] In this work, we use an in-house implementation of the method that has been previously used to demonstrate superior performance of higher order harmonic imaging and design tissue harmonic minimizing pulses for contrast ultrasound.[31-32]

We refer to this simulation model as Virtual Reference Phantom (VRP). No depth or frequency compensation is applied to the returned pressure in this model. Backscatter of the VRP is set as 1. We posit that a VRP may be used in lieu of an actual Reference Phantom scan in a clinical setting for two reasons (a) the single line of echo data generated by angular spectrum is wideband and for a virtual liver-like medium would embody the significant propagative effects of liver tissue such as sound speed, attenuation and non-linearity (b) it would provide a stable reference standard since BSC is set as 1.

The intensity $I_a$ of the received echo pressure $P_a$ at the circular aperture face of the VRP is given by:

$$I_a = \frac{P_a^2}{2\rho c} \qquad (13)$$

ρ is the density of the medium and c is the sound speed. The on-axis intensity $I_{VR}$ of the received echo is considered equivalent to IQ data from the VRP. When used as a reference with an actual scan, the generalized model equation (9) takes the form:

$$\frac{I_S(\omega_0,z_2)-I_S(\omega_0,z_1)}{I_{VR}(\omega_0,z_2)-I_{VR}(\omega_0,z_1)} =$$
$$BSC_S(\omega_0)P(\omega_0)\left[\frac{G_s(\omega_0,z_2)\exp(-4\alpha_f\omega_0\delta)\exp(4\alpha_S\omega_0\delta)\exp(-4\alpha_S\omega_0 z_2)-G_s(\omega_0,z_1)\exp(-4\alpha_f\omega_0 z_1)}{G_{VR}(\omega_0,z_2)\exp(-4\alpha_{VR}\omega_0 z_2)-G_{VR}(\omega_0,z_1)\exp(-4\alpha_{VR}\omega_0 z_1)}\right]$$

(14)

AC of the VRP is $\alpha_{VR}$. $BSC_{VR} = 1$ and $G_{VR} = 1$. Since the beam shape and acoustic power from the actual imaging probe and the circular aperture used in VRP can longer be assumed to be the same, we use a location independent parameter $P(\omega_0)$ to describe the ratio between these parameters for the Sample and the Reference.

We also assume that the transmit focal length and the transmit frequency are the two main characteristics of the ultrasound beam that influence the significant propagative effects that are utilized by the two-parameter model. Equations (10) and (11) take the following form for a VRP.

$$\frac{I_S(\omega_0,z_2)\exp(4(\alpha_f-\alpha_R)\omega_0\delta)-I_S(\omega_0,z_1)}{I_{VR}(\omega_0,z_2)-I_{VR}(\omega_0,z_1)} = BSC_S(\omega_0)P(\omega_0)G_s(\omega_0,z_2)\left[\frac{\exp(-4\alpha_S\omega_0(z_2-\delta))}{\exp(-4\alpha_{VR}\omega_0(z_2-\delta))}\right] \cdot H$$

(15)

$$H = \left[\frac{1-(G_s(\omega_0,z_1)/G_s(\omega_0,z_2))\exp(4\alpha_S\omega_0(z_2-\delta))\exp(-4\alpha_f\omega_0 z_1)\exp(4\alpha_{VR}\omega_0\delta)}{1-\exp(4\alpha_{VR}\omega_0 z_0)}\right]$$

(16)

For values of $\alpha_{VR}$ representative of biological tissue and typical imaging protocols, H can be approximated to $exp(-4\alpha_{VR}\omega_0 z_0)$ when $\alpha_f$ and $\alpha_S$ are not very different from $\alpha_R$ or when $\delta = 0$.

In equation (15), the location independent terms $BSC_S$ and $P$ can be combined into a consolidated backscatter coefficient $BSC_{mea}$. The gain term $G_s(\omega_0,z_2)$ may be approximated as $exp(4\omega_0 \alpha_G z_2)$ where $\alpha_G$ is a coefficient of frequency $\omega_0$ and depth $z_2$ dependent attenuation as applied to a specific imaging protocol. All the exponential terms that are location dependent can then be combined into a consolidated attenuation coefficient $\alpha_{mea}$ in the following equation:

$$\left|\frac{I_S(\omega_0,z_2)\exp(4(\alpha_f-\alpha_R)\omega_0\delta)-I_S(\omega_0,z_1)}{I_{VR}(\omega_0,z_2)-I_{VR}(\omega_0,z_1)}\right| = BSC_{mea}(\omega_0)exp(-4\omega_0\alpha_{mea}(z_2-\delta))$$

(17)

$$\alpha_{mea} = \alpha_S - \alpha_{VR} + \alpha_{VR}(z_0/(z_2-\delta)) - \alpha_G \cdot z_2/(z_2-\delta)$$

(18)

Equation (17) can be log-transformed and linearly regressed to estimate $BSC_{mea}$ and $\alpha_{mea}$. The difference between $\alpha_{mea}$ and $\alpha_S$, and, $BSC_{mea}$ and $BSC_S$ of a homogeneous TM phantom with known AC ($\alpha_S$) and BSC ($BSC_S$) can used to calibrate the VRP for a fixed imaging protocol with a specific imaging probe. A Matlab implementation of this procedure will be adopted in the feasibility study described in Section 5.

## 4. Simulation Study

Ultrasound received echo is simulated using our in-house implementation of nonlinear propagation of ultrasound in tissue based on the angular spectrum approach as mentioned in section 3.2.

### 4.1 Accuracy and Precision of AC and BSC estimates for Matched Sample and Reference

On-axis received echo intensities of full bandwidth of 1.87 MHz centered at the second harmonic (4 MHz) for propagation in a liver-like[30] medium from an apodized 25 mm diameter circular aperture excited with a 2 MHz transmit Gaussian pulse of 5 watt with a transmit focus of 5.5 cm and dynamic focus on receive is used in the estimation of AC and BSC. All tissue characteristics except attenuation coefficient are matched for the Reference and the Sample medium. Sound speed = 1570 m/s, density = 1050 kg/m$^3$, power law index of attenuation = 1.3, nonlinearity parameter ($\beta$) = 4.7, BSC = 1. Attenuation coefficients are set at 0.4 dB/cm-MHz and 0.61 dB/cm-MHz. These values are approximately matched to the average of normal and average of fatty liver values at 3 MHz in Figure 5(a) of Lu et al.[3] With normal liver as Reference, a Matlab implementation of log transform of equation (8) followed by a linear regression computes AC and BSC from the pressure signal intensity of the fatty liver. The proximal ROI for the model is placed at 0.5 cm from the aperture; the distal ROI is placed from axial depths of 2-14 cm; the ROI sizes are 1.5 cm. Pressure values at all frequencies in the band are coherently summed at each axial depth used in the calculation. As shown in figures 3(a) and 3(b), AC and BSC values show remarkable uniformity beyond the focus. AC is recovered to within 5% and BSC is recovered to within 4% accuracy.

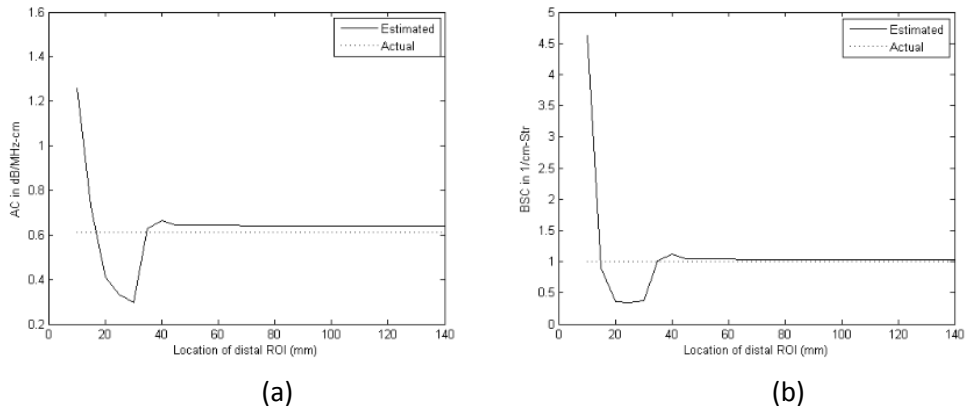

(a)                  (b)

Figure 3. Parameter estimates over axial depth from simulated echoes over a finite bandwidth centered at the second harmonic frequency (4 MHz) based on the new two-parameter model (equation 8). Reference and Sample medium are both 'liver' like and homogeneous. Transmit focus is at 5.5 cm. (a) AC (b) BSC.

### 4.2 Accuracy and Precision of AC and BSC estimates for Unmatched Sample and Reference

Echo data is simulated for a transmit frequency of 2 MHz, transmit focus of 5 cm propagated in a normal liver like medium from a circular aperture of 25 mm with an acoustic power of 5.0 watts. Sound speed = 1570 m/s, density = 1050 kg/m$^3$, power law index of attenuation = 1.3, nonlinearity parameter ($\beta$) = 4.7, AC = 0.4 dB/cm-MHz, BSC = 1. This constitutes the Reference data. There are five Samples of simulated echoes with all parameters unchanged except (1) acoustic power = 10 watts (2) aperture size 40 mm (3) AC = 0.61 dB/cm-MHz (for fatty liver) (4) AC = 0.61 dB/cm-MHz, acoustic power = 10 watts (5) AC = 0.61 dB/cm-MHz, aperture size 40 mm. The estimates of AC and BSC for the five Samples from the solution of equation (8) are shown in figures 4(a) and 4(b) respectively. When the aperture size is unmatched, the estimated AC is lower by ~0.025 dB/cm and estimated BSC is 40% of that of the matched aperture When the acoustic power is unmatched, the estimated AC is higher by ~0.025 dB/cm and estimated BSC is 200% compared to the matched aperture. The shifts remain the same throughout the post-focal region and similar in magnitude for normal and fatty liver.

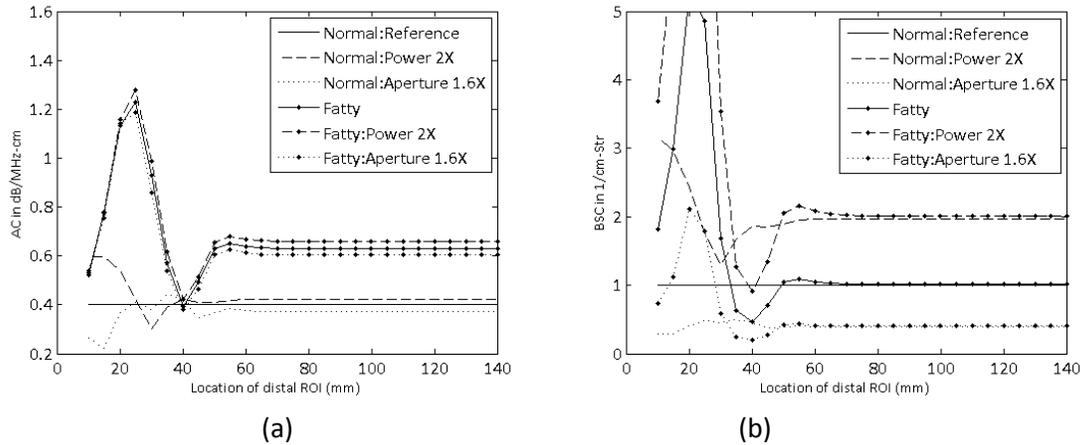

(a) (b)

Figure 4. Parameter estimates over axial depth from simulated echoes over a finite bandwidth centered at the second harmonic frequency (4 MHz). Transmit focus is at 5 cm. Reference and Sample medium are both 'liver' like and homogeneous. The Reference echo is simulated for normal liver from aperture diameter = 25 mm and power = 5 watts. The Sample liver is fatty and echoes are simulated for two aperture diameters - 25 mm and 40 mm and two power settings 5 watts and 10 watts. (a) Estimated AC (b) Estimated BSC.

We next see the effect of unmatched power law index of attenuation between the Sample and the Reference. In general, attenuation of a medium is given by $\alpha = \alpha_0 \, f^n$ where $\alpha_0$ is the coefficient of attenuation, f is the frequency and n is the index.[25] RPM assumes that n=1 for both the Reference and the Sample.[15] Echo data is simulated for a transmit frequency of 2 MHz, transmit focus of 5 cm propagated in a normal liver from a circular aperture of 25 mm with an acoustic power of 5.0 watts. Reference harmonic is computed with sound speed = 1570 m/s, density = 1050 kg/m$^3$, power law index of attenuation = 1.0, nonlinearity parameter ($\beta$) = 2, and attenuation coefficient = 0.49 dB/cm-MHz. Sample harmonic is computed for the same sound speed, density and BSC but power law indices of attenuation = 1.0 and 1.3, nonlinearity parameter ($\beta$) = 4.7 and AC = 0.4 dB/cm-MHz, BSC = 1. AC and BSC estimated from simulated data with matched set at n = 1.0 for the Reference and the Sample and unmatched set (with n = 1.3 for the Sample and n=1.0 for the Reference) are shown in figures 5(a) and 5(b).

The estimated AC in the unmatched case is significantly higher (0.56 dB/cm-MHz) than the actual value (0.4 dB/cm-MHz) in the Sample. In the matched case, error in estimated AC is within 5%. BSC value is within ~5% of the actual value 1 in both cases. As expected, BSC is not too influenced by the mismatch in the power law of attenuation.

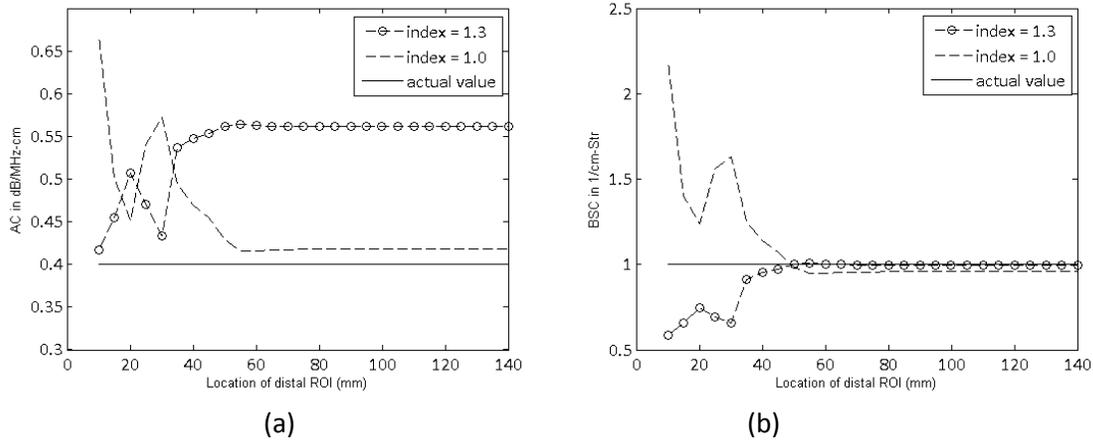

Figure 5. Parameter estimates over axial depth from simulated echoes over a finite bandwidth centered at the second harmonic frequency (4 MHz). Transmit focus is at 5 cm. Simulated echoes are with Reference AC = 0.49 dB/cm-MHz, attenuation index = 1.0 and Sample AC = 0.4 dB/cm-MHz, attenuation index = 1.0 and 1.3. BSC = 1. (a) Estimated AC (b) Estimated BSC.

5. Feasibility Study

Non-alcoholic fatty liver disease (NAFLD) is estimated to occur in 15-30% of the general population and is the most prevalent liver disease.[33] NAFLD could occur as simple steatosis (fatty liver) or non-alcoholic steatohepatitis (NASH) characterized by liver cell injury and inflammation in 10-30% cases. 25-40% cases of NASH patients could progress to advanced fibrosis and 20-30% could eventually end up with cirrhosis. 70-80% people with central obesity have shown evidence of NAFLD on imaging.[34] Biopsy remains the gold standard to diagnose liver fat and liver fibrosis. Biopsy is invasive, prone to read errors, samples only ~1/50,000[th] of liver, and cannot be employed for a screening population.[35] Fibrosis if detected at an early stage can be reversed with medication and changes in life style.

Ultrasound is the most widely used affordable non-invasive imaging tool for distinguishing between normal and fatty liver. Radiologists use perceptual cues such as echogenicity of liver compared to the kidney, size of the right lobe of liver, the sharpness of the diaphragm and vessel walls to categorize liver fat content into mild, moderate and high.[36-37] Modalities used in the evaluation of liver fibrosis within the clinical practice today include Magnetic Resonance Elastography (MRE), Transient Elastography (TE), Acoustic Radiation Force Imaging (ARFI) and Shear Wave Elastography (SWE). Fibroscan, a device that uses TE to integrate the quantification of elasticity with attenuation of shear wave reports two scores - a liver stiffness measurement (LSM) and a Computed Attenuation Parameter (CAP). The data is typically acquired from a representative region within the liver that generates adequate shear wave signal. Fibroscan is simple to use and popular with Hepatology departments across the world.[38-39]

Ultrasound is more accessible and less expensive compared to MRI. However, sonographic assessment of steatosis tends to be subjective, operator dependent and has low reproducibility. Fibroscan has been established to be very effective at quantitatively distinguishing between normal and high steatosis and normal and advanced fibrosis. However, cut-offs between low and moderate states of steatosis, and, low and intermediate levels of fibrosis are more blurred in case of obesity and in the presence of NAFLD. [38-42]

Thus, there continues to exists a clinical need for effective non-invasive screening for patients at risk for NAFLD and monitoring of NAFLD patients for evidence of NASH and fibrosis.[39] A fully automated ultrasound based tool that can estimate acoustic properties of tissue known to alter in the presence of fat; visualize and quantitatively characterize the degree of fat over a larger section of liver than sampled by biopsy with minimal modification in the workflow associated with the standard of care could be very useful for both the Radiologist and the Hepatologist.

In this section, we apply the generalized model described earlier to liver tissue characterization with in-vivo ultrasound harmonic imaging. We outline the clinical study, patient enrolment and attributes; provide system and protocol specification used in the data collection; methods used for calibration, analysis of the estimated parameters AC and BSC and the consolidated score; and sum-up with a demonstration of the results.

5.1 Study Goal and Patient Recruitment

The feasibility study is aimed at evaluating the ability of the generalized two-parameter algorithm for tissue characterization to distinguish between normal and fatty livers for patients with progressive liver disease with NAFLD etiology. Ground truth for the evaluation is provided by liver biopsy (when available) or classification of normal v's fatty liver with fat grade on B-mode ultrasound.

The study is being conducted at the Institute of Liver Biliary Sciences (ILBS), New Delhi, India and approved by the Ethics Committee of ILBS. All patients recruited for the Study have provided informed consent.

Result for 16 subjects have been analyzed in this study. The age of the cohort ranged from 26-58 years except for one subject aged 10-years. There were 2 females and 14 males. B-mode liver ultrasound was performed for all subjects. 5 were found to be normal and constituted the control group (including the 10-year old subject). 4 were found to be fatty (grades ranging from 1-3). 7 subjects underwent liver biopsy and reported varying scores of fatty infiltration and fibrosis. [35,43] Care was taken to make sure that unwilling patients, pregnant women, patients with alcoholic liver disease or history of alcoholism, aged patients, those with gross ascites, poor echo, viral hepatitis or cryptogenic liver disease were not included in this study.

Subject-wise ground truth are as follows: With ultrasound - 1-5: Normal, 6: Fatty grade 2/3, 7: Hepatomegaly Fatty Grade 2, 8: Fatty Grade 2, 9: Fatty Grade 1. With liver biopsy - 10: NAS CRn 2 F0, 11: NAS CRn 5 F1A, 12: NAS CRn 4 F1C, 13: NAS CRn 4 F4, 14: NAS CRn 4 F1A, 15: NAS CRn 5 F2, 16: NAS CRn 5 F2.

5.2 System and Protocol

A GE Logiq E9 Ultrasound Scanner R6 Version with shear wave elastography and the option to extract IQ data is used in the Study. Subjects are scanned with C1-6, a convex probe suitable for abdominal examinations.

A fixed protocol irrespective of the patient habitus or subtype is used in the study. 2-D section of liver is acquired by placing the probe between the intercostal spaces of the right lobe of the liver. It is desirable to acquire a field of view within the liver capsule consisting of mostly parenchyma with minimal vessels and extra-hepatic neighboring anatomies. The operators are assigned randomly; however, they are all experienced sonographers.

B-mode acquisition is performed with harmonic (CHI) with frequency set at 4 MHz, transmit focus of 5 cm for an overall imaging depth of 12 cm. Gain is set at a value of 55 and dynamic range at 66. TGC pods are kept centered. A minimum of three acquisitions are done. 2-5 frames of IQ data are captured for each acquisition and is transferred to a computer using a USB device. Thereafter it is used for offline processing.

5.3 VRP Calibration

The VRP is a single line of echo for propagation in a liver-like medium (AC = 0.4 dB/cm-MHz, sound speed = 1570 m/s, density = 1050 kg/m$^3$, power law index of attenuation = 1.3, nonlinearity parameter ($\beta$) = 4.7 and BSC set at 1) from a 25 mm circular aperture and a 2 MHz centered Gaussian apodized pulse.

A homogeneous liver tissue mimicking (TM) phantom is used in the calibration of the VRP with respect to the C1-6 probe and the imaging protocol used in the study.[44-45] The phantom (Serial No. E2748-1) is housed in a PVC cylinder with an inside diameter of 10 mm and a height of 120 mm. It is filled with a 100 mm layer of Zerdine® gel, followed by a 2 cm layer of acoustic absorbing material. The phantom has the following characteristics: sound speed = 1570 m/s, attenuation coefficient = 0.49 dB/cm-MHz, contrast = 0.4 dB, elasticity = 5.5 kPa. The BSC is nominally set at $3.8 \times 10^{-4}$ /Sr-cm.

Two operators scan the phantom with the imaging protocol nine times each. They position the probe at the center of the top surface of the phantom and capture a minimum of five frames of IQ data for each scan. The data is transferred to a computer for post processing at GE's Research Center at Bengaluru, India.

The size of each frame of IQ data is 975 depth points X 339 scan lines. With VRP as the reference and the phantom as the Sample, AC and BSC for the TM phantom are estimated for each frame of IQ data from equation (17). The proximal ROI is placed at 0.5 cm. The distal ROI's used in the analysis range from scanlines 115-225 and depth points of 367-694 that correspond to the central section of the lateral field of view and a 4 cm depth ranging from 4.5 – 8.5 cm. The choice of the region of parameter estimation is influenced by the need to cover adequate area and at the same time avoid the artifacts near the side and back walls of the phantom. The size of the ROI is set at 0.75 cm in depth and 19 scan lines in width. Example of a frame of the scanned phantom along with the estimated AC and BSC map with VRP is shown in figure 6.

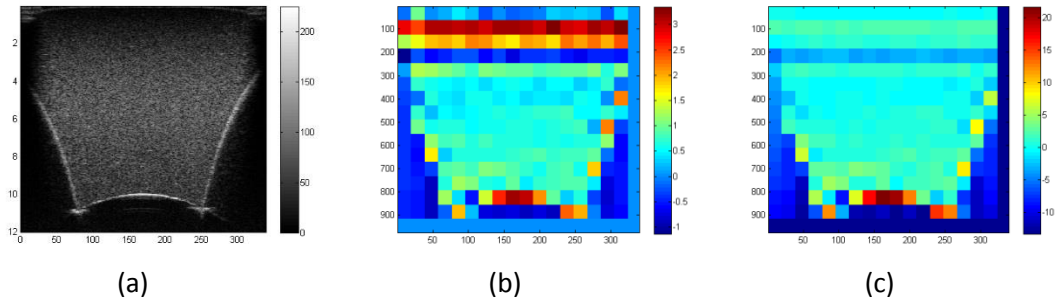

(a) (b) (c)

Figure 6. (a) Example scan of TM Phantom (b) AC map and (c) BSC map estimated from the application of equation (17) with VRP as Reference.

The median of the AC and BSC distribution from the region of analysis in each frame is representative of the field of view associated with the frame. AC and BSC estimates from the frames of a subset of four acquisitions (20 values) for one operator, and three acquisitions (15 values) for the other operator are found to be normally distributed (p-values > 0.05 in Anderson Darling normality test). The mean values of AC $\alpha_{mea}$ measured for operator 1 and 2 are: 0.56 ± 0.004 dB/cm-MHz and 0.62 ± 0.04 dB/cm-MHz respectively. The mean values of BSC $BSC_{mea}$ (expressed in base 10 logarithm) for operator 1 and 2 are: 0.11 ± 0.05 and 0.5 ± 0.2 respectively.

VRP bias is calculated following the method of calibration laid out in section 3.2. Therefore, $\alpha_{bias} = \alpha_{mea} - \alpha_S$ and $BSC_{bias} = BSC_{mea} - BSC_S$. Values of the Sample are from the TM phantom $\alpha_S$ = 0.49 dB/cm-MHz and log10($BSC_S$) = -3.42. The measured values of $\alpha_{mea}$ from operator 1 and $BSC_{mea}$ from operator 2 are chosen to compute the VRP bias since the errors associated with those measurements are lower in comparison with those of the other operator. Thus $\alpha_{bias}$ = 0.07 dB/cm-MHz and log10($BSC_{bias}$)= 3.92.

VRP constitutes the Reference scan and the bias values calculated here are used throughout the parameter estimates $\alpha_S$ and $BSC_S$ in the actual liver scans.

5.4 Application of the Generalized Model

Equation (17) is used in the estimation of AC and BSC at every data point in the post-focal region from the IQ data captured from scans of each of the 16 subjects in the Study cohort. The second harmonic signal is extracted from the wideband received IQ at the harmonic (CHI) frequency 4 MHz. Width of the bandpass filter is set at 3.25 MHz and the sampling frequency is 10 MHz.

A set of automation methods are implemented in Matlab to further simplify the application of the generalized model in actual practice.

In the first of these we automate the placement of the proximal ROI from an image frame based on its closeness to the probe and apply a homogeneity criterion to be consistent with the assumptions of the RPM.[11] Given the imaging protocol, a 0.75 cm ROI is large enough to capture adequate information from the signal to guarantee a statistically acceptable linear regression. If the ROI is too large, there is more chance for it to contain neighboring structures and high intensity interfaces often present in the subcutaneous layer that could lead to reduced accuracy of regression. We employ the gamma distribution for the selection of proximal ROI.[46] The two-parameter gamma probability distribution $P(x)$ with shape $\alpha$ and scale $\theta$ is given by:

$$P(x) = \frac{x^{\alpha-1}.e^{-x/\theta}}{\Gamma(\alpha).\theta^\alpha} \qquad (19)$$

The shape parameter of the gamma distribution is a simple measure of heterogeneity. Generally, homogenous regions have larger shape parameters. When a neighboring structure is additionally captured within the ROI then the shape parameter reduces. This relationship can be exploited to select the location of the proximal ROI by dividing the IQ data into overlapping ROIs of axial size $r$ and $n$ scan lines. On each rectangular ROI, a gamma distribution can be fitted and its parameters are calculated. In our case, we compare shape parameters calculated from multiple ROI's of axial size 0.75 cm and 19 scanlines over a ROI start depth between the range of 0.6 – 1.2 mm and lateral shift varying between -50 to +50 of the central scan line in the image frame. The proximal ROI location is determined by the ROI start depth with the largest shape parameter within this search range.

The IQ distribution in the proximal ROI tends to be noisy and the low second harmonic signal in the near-field often in the 'obese' subcutaneous layer that has many interfaces is amplified by the gain term. A Gaussian smoothing (window size = 5 samples) is applied within the IQ distribution in the ROI along the depth direction to remove outliers, which improves the accuracy of the linear fit performed for parameter estimation using the two-parameter model.

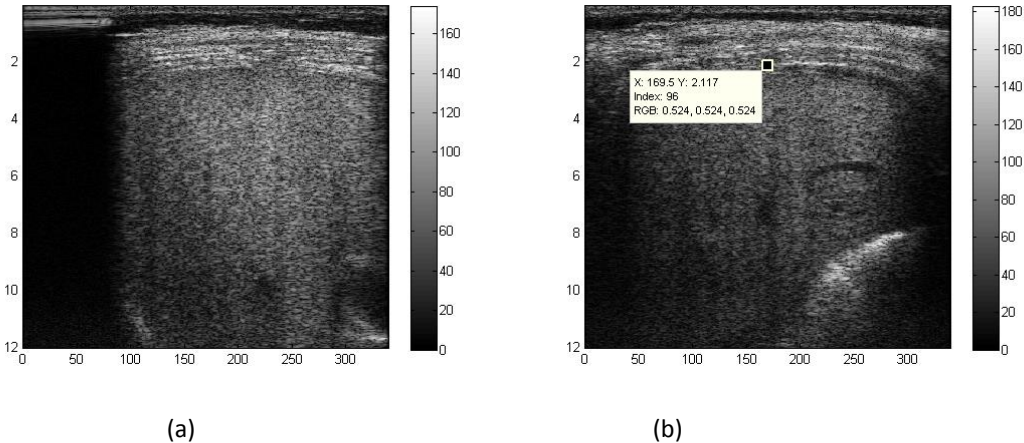

(a)          (b)

Figure 7. (a) Example scan of liver with dark lines on the left symptomatic of lack of gel contact (b) manual extraction of thickness of subcutaneous layer.

Lack of adequate probe contact due to ribs in intercostal liver scans often leads to darkening of the image as illustrated in figure 7(a). A simple Otsu-threshold is applied on the acquired IQ data prior to parameter estimation so that estimates are primarily from parenchymal regions only.[47]

The thickness of the layer of subcutaneous fat is extracted manually from the central scan line in the field of view by visually inspecting of the interface between the layer and the underlying liver as shown in figure 7(b). In solving equation (17) with finite $\delta$, we assume a value of 1.05 dB/cm-MHz for $\alpha_f$. This is within the range for muscle (1.3 dB/cm-MHz) and fat (0.6 dB/cm-MHz) reported by Goss et al.[48] and compares closely with a value 1 dB/cm-MHz used in quantification of liver fat in recent literature.[5]

5.5 Data Analysis

AC and BSC are estimated at every IQ sample. Non-overlapping ROI's (axial size = 0.75 cm and 19 scanlines) post-focus (>5.0 cm) are considered in the evaluation of a representative AC and BSC for the liver parenchymal region in the field-of-view once they are validated through checks to ensure that the data meets a few basic quality criteria. The checks consist of an outlier rejection strategy and distribution skewness threshold. ROI's in which >10% AC values are negative are excluded. Negative AC values are physically unacceptable. ROI's in which distributions of AC or BSC are very skewed are also rejected. The ratio of interquartile range/median should be low for a tight distribution with low variability. For normal distribution 66% data is within mean - SD and mean + SD. Here we check that (Range of Q3-Q1)/Median <=66% in the ROI for both AC and BSC independently and reject the ROI if it is not so. To be close to a normal distribution each quartile should be checked separately. Fibroscan sets this at 30% for each quartile.[41] We set a less conservative bound for both tightness of the distribution and the extent to which it can depart from normality.

Number of ROI's that meet the quality criteria is contingent on both quality of the scan (steady hand and predominantly parenchyma in the field of view) and heterogeneity of tissue at a scale that is discernable by ultrasound at a frequency ranging from ~ 2.5 – 5.5 MHz contained in the harmonic band. The median of AC and BSC ($\alpha_S$ and $BSC_S$) distribution from the acceptable ROI's of each frame are used as a representative measure of these parameters from the field of view. The median values of all the frames across the scans are used to compute a mean and standard deviation of AC and BSC of each subject.

The two parameters $AC_{mean}$ and $BSC_{mean}$ thus calculated are combined to form a dimensionless normalized score mathematically defined as:

$$Liver\ Score = \frac{\sqrt{(AC_{mean}/AC_{ref\_normal})^2 + (BSC_{mean}/BSC_{ref\_normal})^2}}{\sqrt{2}} \qquad (14)$$

$AC_{ref\_normal}$ and $BSC_{ref\_normal}$ are the reference values for normal liver tissue. We have taken these values to be 0.46 dB/cm-MHz and 3.8 X 10$^{-4}$ /Sr-cm respectively. These lie in the range of values reported in literature. [3,29,49] The Liver Score for normal liver is thus expected to be ~ 1, and increase as liver becomes fatty.

5.6 Results

Number of scans, total number of frames over which analysis is performed and mean number of accepted ROI's, mean and standard deviation of AC and BSC for each subject is shown in Table 1. As mentioned in section 5.1, Subjects 1-5 have normal liver on ultrasound, 6-9 show fatty liver on ultrasound, 10-16 are patients with biopsy and varying degrees of fat and fibrosis.

Table 1. Subject-wise scans and analysis

| Subject# | # scans | Total # frames | Mean # ROI's | SD | Mean AC (dB/cm-MHz) | SD | Mean BSC (log10) | SD |
|---|---|---|---|---|---|---|---|---|
| 1 | 3 | 9 | 55 | 11 | 0.51 | 0.09 | -3.52 | 0.34 |
| 2 | 2 | 10 | 54 | 10 | 0.34 | 0.01 | -3.72 | 0.04 |
| 3 | 4 | 16 | 28 | 5 | 0.45 | 0.05 | -3.99 | 0.31 |
| 4 | 3 | 14 | 15 | 13 | 0.43 | 0.15 | -3.82 | 0.28 |
| 5 | 4 | 20 | 16 | 11 | 0.47 | 0.12 | -4.01 | 0.26 |
| 6 | 4 | 16 | 86 | 14 | 0.56 | 0.15 | -2.91 | 0.09 |
| 7 | 4 | 17 | 127 | 11 | 0.41 | 0.04 | -2.97 | 0.29 |
| 8 | 4 | 19 | 112 | 32 | 0.39 | 0.01 | -3.14 | 0.18 |
| 9 | 4 | 20 | 38 | 11 | 0.48 | 0.05 | -3.64 | 0.35 |
| 10 | 5 | 17 | 56 | 6 | 0.59 | 0.03 | -3.08 | 0.10 |
| 11 | 4 | 20 | 146 | 5 | 0.43 | 0.01 | -2.92 | 0.04 |
| 12 | 3 | 12 | 107 | 5 | 0.34 | 0.01 | -3.26 | 0.04 |
| 13 | 4 | 16 | 52 | 9 | 0.44 | 0.07 | -3.30 | 0.34 |
| 14 | 4 | 18 | 66 | 12 | 0.38 | 0.03 | -3.72 | 0.12 |
| 15 | 4 | 18 | 73 | 8 | 0.49 | 0.02 | -3.20 | 0.12 |
| 16 | 4 | 16 | 50 | 13 | 0.71 | 0.04 | -2.60 | 0.14 |

Mean number of acceptable ROI's vary across the subjects. High standard deviations (SD) in the number of acceptable ROI's are owing to very different field of views for each scan. Since operators were assigned at random, there is some variability in the quality of acquisitions. Mean AC of normal subjects and those with progressive NAFLD are 0.44 ± 0.08 and 0.48 ± 0.09 dB/cm-MHz respectively. Mean BSC of normal subjects and those with progressive NAFLD are -3.81 ± 0.25 and -3.16 ± 0.2 in log(10) scale respectively.

AC and BSC map for a single frame of acquisition for an obese subject with normal liver is shown in figure 8. AC and BSC map for an obese subject with fatty liver is shown in figure 9. AC and BSC map for an obese subject with fatty infiltration and fibrosis stage 2 is shown in figure 10.

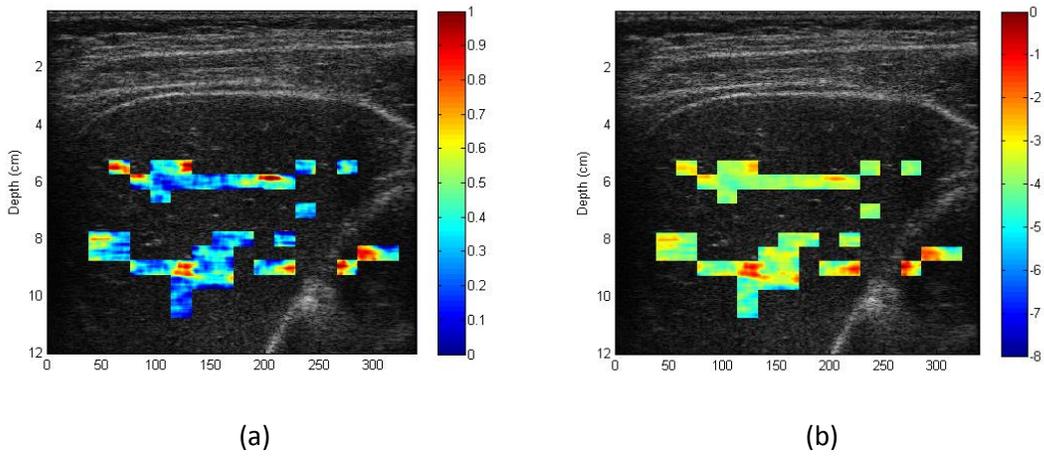

Figure 8. Subject#2 with normal liver on ultrasound (a) AC map (b) BSC map.

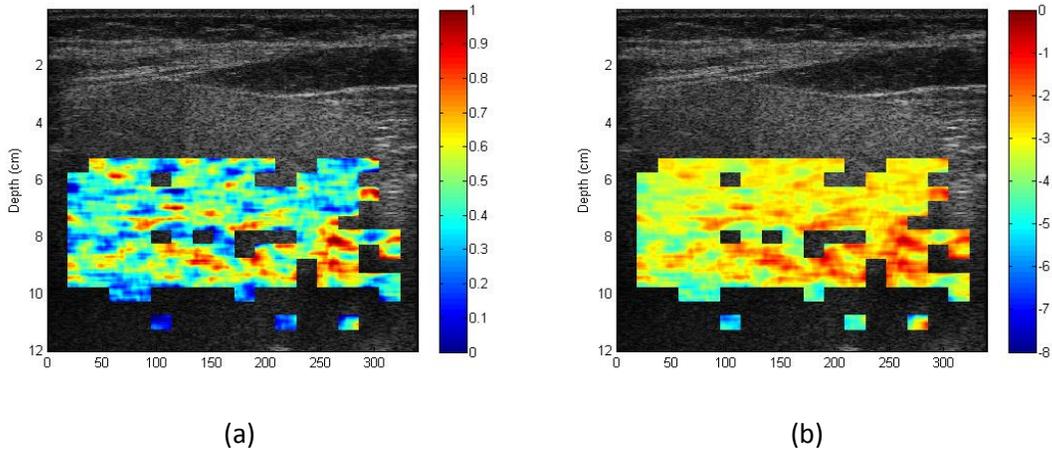

Figure 9. Subject#6 with fatty liver grade 2/3 on ultrasound (a) AC map (b) BSC map.

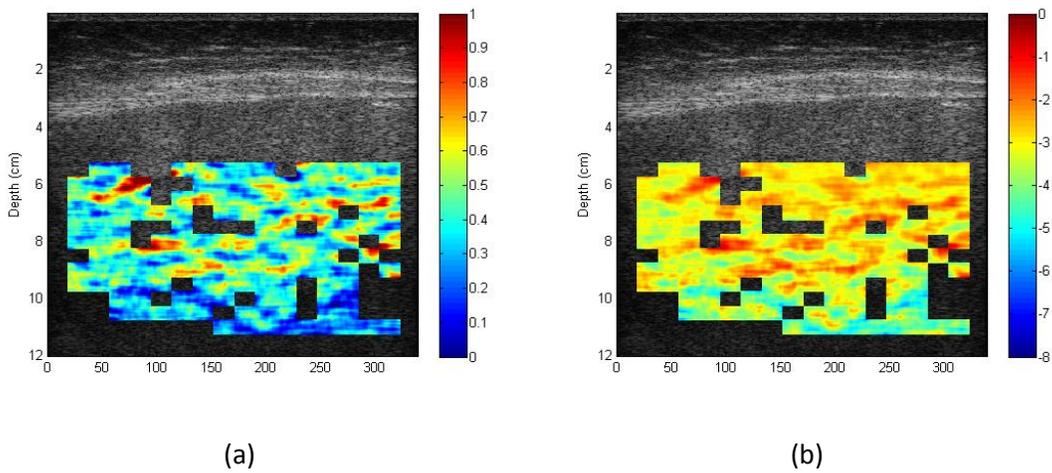

Figure 10. Subject#11 with NASH on biopsy (a) AC map (b) BSC map.

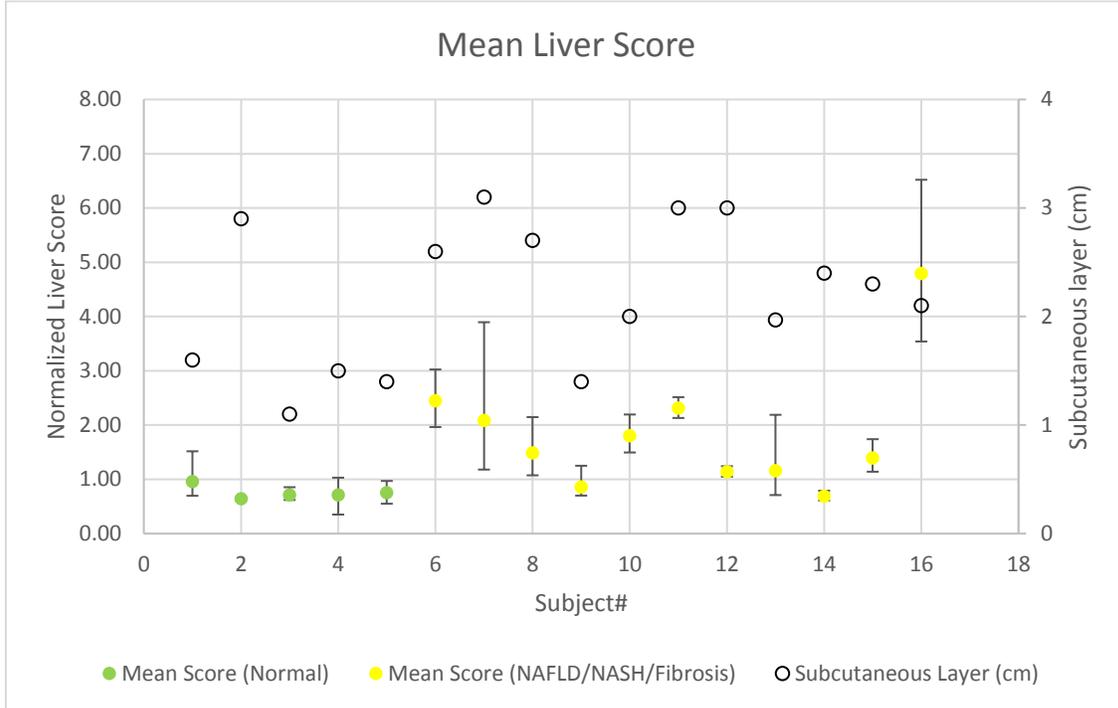

Figure 11. Mean scores (primary vertical axis) for liver tissue characterization of 16 subjects. Error bars are a consolidated indicator of tissue heterogeneity and quality of scans. Subcutaneous fat layer (secondary vertical axis) for each subject is also shown.

Liver scores with errors bars calculated using mean ± standard deviation of AC and BSC values are shown graphically for all subjects in figure 11. The thickness of the subcutaneous or 'obese' layer is also plotted on the same chart. In the subject cohort that was studied, the thickness of this layer varied from 1.1 – 3.1 cm. Subjects with progressive NAFLD are in general more obese than normal subjects (layer thickness 2.4 ± 0.5 cm compared to 1.7 ± 0.7 cm).

The mean liver scores are ≤1 for normal liver and >1 for NAFLD (fat grade on ultrasound ≥ 1, NASH and Fibrosis). There are two exceptions: Ultrasound for Subject#9 was rated as fat grade 1 by the Radiologist but its liver score is <1. Subject#14 was rated as NASH by biopsy but its liver score is <1.

The estimation of mean AC, BSC and the corresponding score can be affected considerably by variability in the field of view of the scans for the same subject. In figure 12, we show the BSC maps for two different scans on subject# 9. Median BSC in the first case is lower while the value in the second case is higher. The estimation can also be impacted when the ultrasound beam encounters strong attenuators like ribs in the near field. This is evident in the rib shadows in the scans for subject#14 illustrated by a frame and the corresponding BSC map in figure 13.

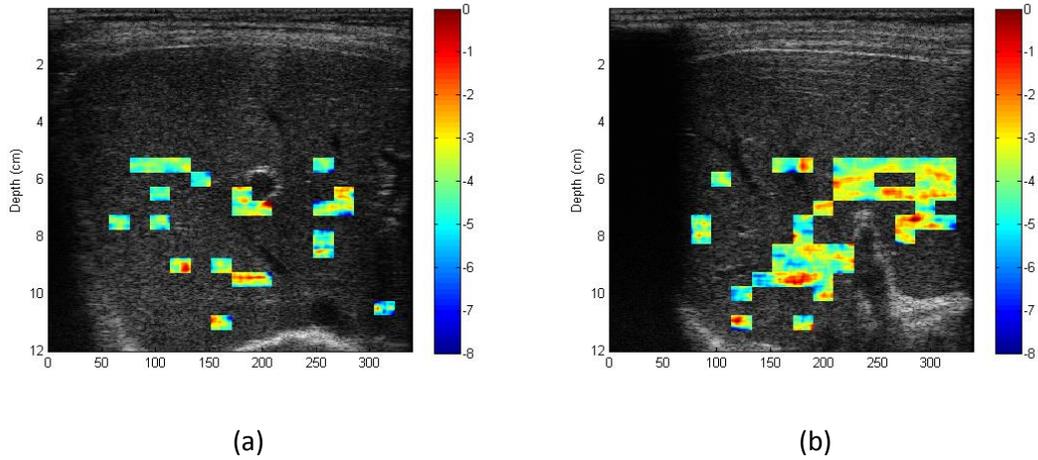

(a)　　　　　　　　　　　　　　　(b)

Figure 12. BSC maps in Subject#9 (a) First scan (b) Third scan. Field of views are different. Median BSC is higher is the third scan.

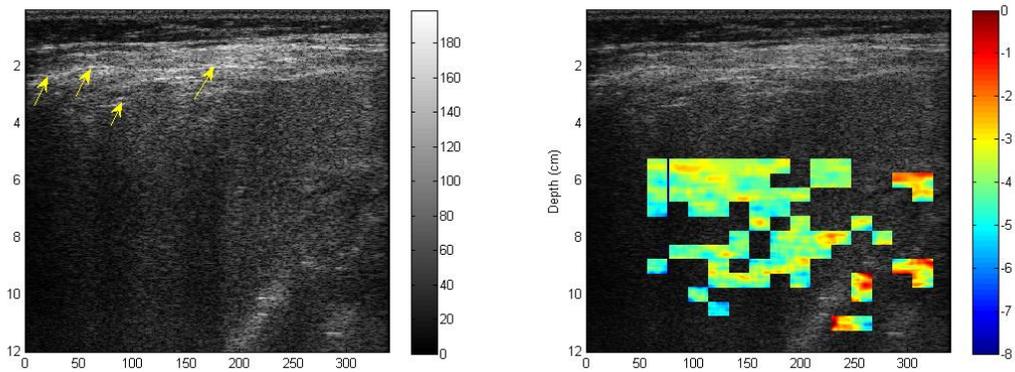

Figure 13. Subject#14 (a) Wideband image shows ribs in the near field (b) BSC map in the corresponding region. BSC is underestimated due to inadequate compensation for attenuation in the subcutaneous layer.

6. Conclusions

In this paper, we have proposed a new two-parameter model for the estimation of AC and BSC for tissue characterization with harmonic imaging using a VRP. We have applied the method to simulated echo intensities and shown that it can recover AC and BSC simultaneously with accuracy and precision over a post-focal depth of tissue in homogeneous media.

We have also formulated a generalized version of the model applicable to commercial scanners to account for pre-applied depth and frequency gain compensation and during in-vivo imaging to account for subcutaneous layer of skin-muscle-fat that the ultrasound may interrogate before entering the tissue of interest.

In a feasibility study with 16 patients (5 normal and 11 with progressive liver disease with the etiology of NAFLD), we have demonstrated the ability of the method to discriminate between normal and diseased liver in 14/16 cases using a normalized score that combines AC and BSC.

The performance of the algorithm is contingent on data quality and the extent to which the assumptions are valid. Data quality and consistency can be improved through greater standardization of the scanning location and process. 1/16 cases (subject#9) is misclassified as normal due to inconsistency in the scans. The attenuation coefficient of the subcutaneous layer is assumed to be 1.05 dB/cm-MHz. 1/16 cases (subject#14) is incorrectly classified as normal due to rib bone in the near field that causes sharper attenuation than has been assumed. Theoretically, the model is expected to underestimate scores for fatty livers ($\alpha_S > \alpha_R$) as obesity increases. The stress test is then about – can the scores in such livers still distinguish them from normal livers – figure 11 suggests so from the 7/8 obese cases of diseased livers with >2.5 cm of subcutaneous layer.

The higher BSC from livers of 9/11 subjects with varying levels of fatty infiltration with or without the presence of fibrosis in the presence of obesity is in line with the findings of Wallace et al.[22] and Lin et al.[24] and indicates that harmonic backscatter could indeed provide a mechanism for discrimination of early stages of fibrosis in subjects with NAFLD. 4/5 subjects (#s 11, 12, 14, 15 and 16) with fibrosis (F≤2) have mean scores > 1. More extensive clinical studies will be required for further validation of this mechanism and its usefulness in screening and monitoring patients with NAFLD.

The overall algorithm can be further automated and improved by incorporating learning-based methods to identify the subcutaneous layer and parenchymal region to constrain parameter estimation in liver tissue alone.

The generalized model for tissue characterization presented in this paper can also be adapted to estimate AC and BSC for other tissue e.g. breast, ovary with appropriate anatomical and acoustic models for VRP and to account for layered inhomogeneity.

Acknowledgements: Nithin Nagaraj and Nitin Singhal contributed to this work while they were with GE Global Research, Bangalore, India. Kajoli would like to thank Gang Cheng, Xiaodong Han, Prasad Sudhakar, Neha Tademeti, Aditi Kathpalia and Wenting Ye for participating in the intermediate testing and refinement of the models with phantom and volunteer studies (not reported here). She would specially like to thank the support she received from her GE colleagues Ravinder Kaur, Dhanya Nair, Larry Mo, Mike Macdonald, Rimon Tadross, David Becker, Tomohiro Negishi, Marc Barlow, Mike Washburn; and Dr. S.K. Sarin from ILBS.